\documentclass[prl,twocolumn,preprintnumbers,amsmath,amssymb,superscriptaddress]{revtex4}

\usepackage{graphicx}

\begin{document}

\title{Contribution of dielectrics to frequency and noise\\ of NbTiN superconducting resonators}

\author{R. Barends}
\affiliation{Kavli Institute of NanoScience, Faculty of Applied
Sciences, Delft University of Technology, Lorentzweg 1, 2628 CJ
Delft, The Netherlands}

\author{H. L. Hortensius}
\affiliation{Kavli Institute of NanoScience, Faculty of Applied
Sciences, Delft University of Technology, Lorentzweg 1, 2628 CJ
Delft, The Netherlands}

\author{T. Zijlstra}
\affiliation{Kavli Institute of NanoScience, Faculty of Applied
Sciences, Delft University of Technology, Lorentzweg 1, 2628 CJ
Delft, The Netherlands}

\author{J. J. A. Baselmans}
\affiliation{SRON Netherlands Institute for Space Research,
Sorbonnelaan 2, 3584 CA Utrecht, The Netherlands}

\author{S. J. C. Yates}
\affiliation{SRON Netherlands Institute for Space Research,
Sorbonnelaan 2, 3584 CA Utrecht, The Netherlands}

\author{J. R. Gao}
\affiliation{Kavli Institute of NanoScience, Faculty of Applied
Sciences, Delft University of Technology, Lorentzweg 1, 2628 CJ
Delft, The Netherlands}

\affiliation{SRON Netherlands Institute for Space Research,
Sorbonnelaan 2, 3584 CA Utrecht, The Netherlands}

\author{T. M. Klapwijk}
\affiliation{Kavli Institute of NanoScience, Faculty of Applied
Sciences, Delft University of Technology, Lorentzweg 1, 2628 CJ
Delft, The Netherlands}

\date{\today}

\begin{abstract}
We study NbTiN resonators by measurements of the temperature
dependent resonance frequency and frequency noise. Additionally,
resonators are studied covered with SiO$_x$ dielectric layers of
various thicknesses. The resonance frequency develops a
non-monotonic temperature dependence with increasing SiO$_x$ layer
thickness. The increase in the noise is independent of the SiO$_x$
thickness, demonstrating that the noise is not dominantly related to
the low temperature resonance frequency deviations.
\end{abstract}

\maketitle

The interest in the low temperature properties of superconducting
resonators for photon detection \cite{day,lehnert}, quantum
computation \cite{wallraff,palacios} and quasiparticle relaxation
experiments \cite{barends} increases. In principle these properties
are determined by the superconductor, but in practice excess noise
and low temperature deviations in the resonance frequency have been
observed, which are attributed to dielectrics. It is understood that
two-level systems (TLS) in dielectrics in the active region of
resonators contribute to limiting the quality factor and phase
coherence, cause noise and affect the permittivity $\epsilon$
\cite{martinis,phillips,gaoAPL,gaoAPL2}. In order to identify the
physical mechanisms through which two-level systems in dielectrics
affect the microwave properties of superconducting films, we have
chosen to study NbTiN resonators with various coverages of SiO$_x$.
We find that NbTiN follows the Mattis-Bardeen theory for the complex
conductivity more closely than any of the other previously used
superconductors (Nb, Ta and Al) \cite{mattis}. We demonstrate that
deviations from the ideal superconducting properties can be
generated by covering the resonators with a thin amorphous
dielectric layer. In addition, we find that this dielectric layer
affects the noise and the permittivity differently.

\begin{figure}[b]
    \centering
    \includegraphics[width=1\linewidth]{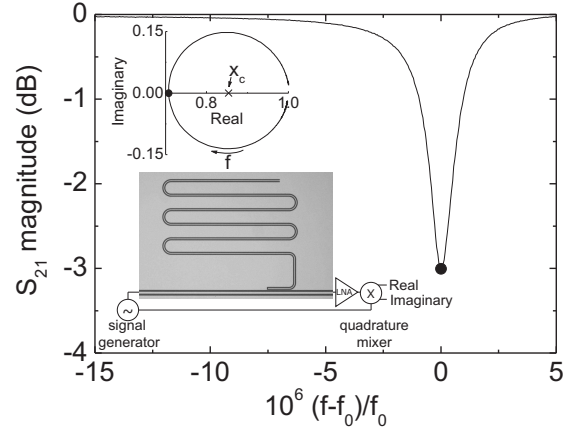}
    \caption{The resonance feature appears as a dip in the magnitude and circle in the complex plane (upper inset) of the feedline transmission $S_{21}$.
    The quarter wavelength resonator is capacitively coupled to a feedline, formed by the superconducting film (gray) interrupted by slits (black) (lower inset).
    The loaded quality factor for this NbTiN resonator is $Q_l=630\cdot 10^3$, its resonance frequency is $f_0=4.47$ GHz.
    The feedline transmission is measured with a signal generator, low noise amplifier (LNA) and quadrature mixer.}
    \label{figure:fig1}
\end{figure}

We have made thin film NbTiN coplanar waveguide (CPW) quarter
wavelength resonators. The resonator (see lower inset Fig.
\ref{figure:fig1}) is formed by a central line, 3 $\mu$m wide, and
slits of 2 $\mu$m wide, with a NbTiN film thickness of 300 nm. The
resonator is capacitively coupled to the feedline by placing the
open end alongside it. The complex conductivity $\sigma_1-i
\sigma_2$, with $\sigma_1$ reflecting the conductivity by
quasiparticles and $\sigma_2$ arising from the accelerative response
of the Cooper pair condensate, leads to a kinetic inductance $L_k
\propto 1/d 2 \pi f \sigma_2$ for thin films with thickness $d$
\cite{mattis,tinkham}. The resonance frequency is controlled by the
kinetic inductance and permittivity,
$f_0=1/4l\sqrt{(L_g+L_k)C(\epsilon)}$, with $l$ the length of the
central line and $L_g$ the geometric inductance and $C \propto
\epsilon$ the capacitance per unit length. The resonance frequency
is therefore a direct probe for both the complex conductivity and
the permittivity,
\begin{equation}
\label{equation:f0}
\frac{\delta f_0}{f_0} = \frac{\alpha}{2} \frac{\delta \sigma_2}{\sigma_2} - \frac{F}{2} \frac{\delta \epsilon}{\epsilon},
\end{equation}
with $\alpha=L_k/(L_g+L_k)$ the kinetic inductance fraction and $F$
a factor which takes into account the active part of the resonator
filled with the dielectric, as argued by Gao \textit{et al.}
\cite{gaoAPL2}. Resonance frequencies lie between 3-6 GHz. Near the
resonance frequency the forward transmission of the feedline
$S_{21}$ shows a dip in the magnitude when measured as a function of
the microwave frequency $f$ (Fig. \ref{figure:fig1}) and traces a
circle in the complex plane (upper inset Fig. \ref{figure:fig1}). In
our experiment we measure both the temperature dependence of $f_0$
as well as the noise in $f_0$ in both bare resonators and resonators
covered with SiO$_x$. The combination of these measurements allows
us to study the possible correlation between the noise and resonance
frequency deviations.

The NbTiN film, 300 nm thick, is deposited by DC magnetron
sputtering on a HF-cleaned high resistivity ($>$1 k$\Omega$cm)
(100)-oriented silicon substrate. Patterning is done using optical
lithography and reactive ion etching in a SF$_6$/O$_2$ plasma. The
critical temperature is $T_c$=14.8 K, the low temperature
resistivity is $\rho$=170 $\mu\Omega$cm and the residual resistance
ratio is 0.94. After patterning we have covered several samples with
a 10, 40 and 160 nm thick SiO$_x$ layer, RF sputtered from a SiO$_2$
target and $x$ is expected to be close to 2. Three chips are partly
covered with SiO$_x$, i.e. each chip contains both fully covered and
uncovered resonators, the latter serving as reference, and a fourth
chip is kept uncovered. Measurements are done using a He-3 sorption
cooler in a cryostat, with the sample space surrounded by a
superconducting magnetic shield. The complex transmission $S_{21}$
is measured by applying a signal along the feedline and amplifying
and mixing it with a copy of the original signal in a quadrature
mixer, whose outputs are proportional to the real and imaginary
parts of $S_{21}$ (lower inset Fig. \ref{figure:fig1}). We find
quality factors in the order of $10^6$.

\begin{figure}[t]
    \centering
    \includegraphics[width=1\linewidth]{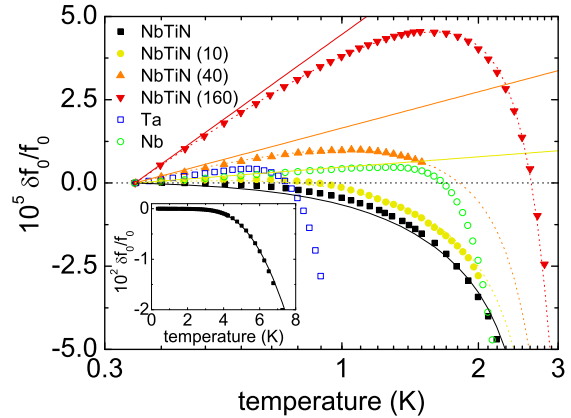}
    \caption{(Color online) The temperature dependence of the resonance frequency of NbTiN samples with no coverage,
    NbTiN samples with a 10 nm, 40 nm or 160 nm thick SiO$_x$ coverage, and samples of Ta and Nb.
    The solid yellow, orange and red lines are fits of the low temperature data to Eq. \ref{equation:permittivity}.
    The inset shows the temperature dependence of the resonance frequency of a NbTiN sample over a broader temperature range
    which closely follows Mattis-Bardeen theory (solid black line) \cite{mattis}.
    The superposition of the Mattis-Bardeen theory (solid black line) and fits to the logarithmic temperature dependence
    found in data of covered samples (solid yellow, orange and red lines) yields the dotted yellow, orange and red lines (Eq. \ref{equation:f0}).}
    \label{figure:fig2}
\end{figure}

The temperature dependence of the resonance frequency is shown in
Fig. \ref{figure:fig2} down to a temperature of 350 mK. The data
shown is representative for all samples. NbTiN (black squares)
closely follows the theoretical expression for the complex
conductivity (black line) \cite{mattis} (inset Fig.
\ref{figure:fig2} and main figure), provided a broadening parameter
of $\Gamma=17$ $\mu$eV is included in the density of states,
following the approach in Ref. \cite{dynes}. We find a kinetic
inductance fraction of $\alpha=0.35$, from which we infer a magnetic
penetration depth of $\lambda=340$ nm \cite{booth}. The resonance
frequency decreases monotonically with increasing bath temperature.
For both 150 nm Ta on Si (blue open squares) ($T_c=4.43$ K) and 100
nm thick Nb on Si (green open circles) ($T_c=9.23$ K), the resonance
frequency increases with increasing temperature at low temperatures,
displaying a non-monotonic temperature dependence over the full
range. Bare NbTiN is in this respect different from Ta and Nb.
However, the NbTiN samples covered with a 10 nm (yellow circles), 40
nm (orange triangles pointing upwards) and 160 nm (red triangles
pointing downwards) SiO$_x$ layer exhibit a non-monotonicity in the
resonance frequency temperature dependence, an effect stronger in
samples with thicker layers.

The data in Fig. \ref{figure:fig2} clearly demonstrate that a
non-monotonic resonance frequency temperature dependence, similarly
to what we find for Ta and Nb samples and for samples of Al on Si
\cite{baselmansjltp}, and what has been reported for Nb on sapphire
samples \cite{gaoAPL2}, can be created in NbTiN by covering the
samples with SiO$_x$. SiO$_x$ is an amorphous dielectric and
contains a large amount of defects \cite{griscom}, giving rise to
two-level systems having a dipole moment, which affect the high
frequency properties \cite{golding,schikfus}. At low temperatures
the resonant interaction of the dipole two-level systems with the
electric fields dominates and leads to a temperature dependent
permittivity (in the limit $kT>hf$) \cite{phillips},
\begin{equation}
\label{equation:permittivity}
\frac{\delta \epsilon}{\epsilon} = - \frac{2 p^2 P}{\epsilon}
\ln \Big(\frac{T}{T_0} \Big),
\end{equation}
with $p$ the dipole moment, $P$ the density of states and $T_0$ an
arbitrary reference temperature (here we choose $T_0$ equal to the
base temperature of 350 mK). At low temperatures the resonance
frequency increases logarithmically with increasing temperature,
indicated by the solid yellow, orange and red lines in Fig.
\ref{figure:fig2}. The slope of the logarithmic increase scales
linearly with the SiO$_x$ thickness. The superposition of the
complex conductivity (solid black line) and the fits to the
logarithmic temperature dependence (Eq. \ref{equation:permittivity})
closely describes the observed resonance frequency (Eq.
\ref{equation:f0}, dotted lines). The logarithmic temperature
dependence and the thickness scaling indicate that dipole two-level
systems distributed in the volume of the SiO$_x$ affect the
permittivity. At higher temperatures the complex conductivity
dominates, leading to a decrease of the resonance frequency.

\begin{figure}[t]
    \centering
    \includegraphics[width=1\linewidth]{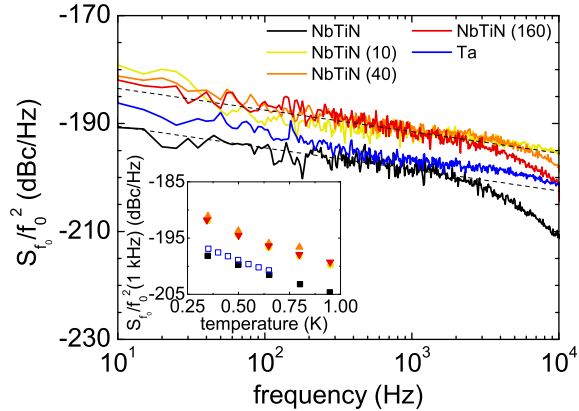}
    \caption{(Color online) Noise spectra of the normalized frequency for NbTiN samples without and with a 10, 40 or 160 nm thick SiO$_x$ layer as well as for Ta.
    The bath temperature is 350 mK and the internal resonator power is $P_{int} \approx -30$ dBm (standing wave amplitude $V_{rms} \approx 14$ mV).
    The dashed lines are fits to the spectral shape, $S_{f_0}/f_0^2 \propto f^{-0.4}$.
    The inset shows the temperature dependence of the noise spectra at 1 kHz (see legend Fig. \ref{figure:fig2}).}
    \label{figure:fig3}
\end{figure}

In the second experiment we have measured the normalized frequency
noise spectra $S_{f_0}/f_0^2$ of bare NbTiN and Ta samples and NbTiN
samples with various SiO$_x$ coverages (Fig. \ref{figure:fig3}). The
noise is measured by converting the complex transmission at the
resonance frequency into a phase $\theta=\arctan
[\mathrm{Im}(S_{21})/(x_c-\mathrm{Re}(S_{21}))] $ with $x_c$ the
midpoint of the resonance circle (see upper inset Fig.
\ref{figure:fig1}). The frequency is related to the phase by:
$\theta=-4 Q_l \frac{\delta f_0}{f_0}$, with $Q_l$ the resonator
loaded quality factor. The power spectral density is calculated by:
$S_{f_0}/f_0^2=S_\theta / (4Q_l)^2$. The noise spectra of samples of
NbTiN (black) and NbTiN with a 10 nm (yellow), 40 nm (orange) and
160 nm (red) thick SiO$_x$ layer follow $S_{f_0}/f_0^2 \propto
f^{-0.4}$ (dashed black) until a roll-off at a frequency in the
order of 10 kHz. The roll-off is due to the resonator-specific
response time and is a function of the loaded quality factor and
resonance frequency. We find that the noise is significantly
increased by approximately 7 dBc/Hz as soon as the samples are
covered by SiO$_x$ and that this increase is independent of the
further increase in SiO$_x$ layer thickness. This behavior persists
with increasing temperature, where the noise decreases (inset Fig.
\ref{figure:fig3}), consistent with recent observations for Nb
\cite{kumar}.

These measurements clearly show that the increase in the noise is
independent of the SiO$_x$ layer thickness, whereas the change in
resonance frequency is thickness dependent. It has recently been
argued, in independent work \cite{gaoAPL2,gaoarXiv}, that the
dielectric influences \emph{both} the resonance frequency and the
noise through the capacitance. In this work we have demonstrated
that indeed the resonance frequency is controlled by the bulk of the
dielectric. However, the observed noise enhancement appears due to
the interface. The latter suggests that it is related to
quasiparticle trapping and release at the interface, influencing the
inductance rather than the capacitance. We find that the noise of
NbTiN samples covered with SiO$_x$ has a spectral shape and
temperature dependence which is very comparable to the noise of
NbTiN samples without coverage and also of Ta samples. In addition,
the noise of NbTiN and Ta samples is very similar, while the
temperature dependence of the resonance frequency is significantly
different. This points towards an interpretation of the noise in
terms of inductance fluctuations.

In summary, we conclude that the frequency noise and the low
temperature deviations in the resonance frequency of planar
superconducting resonators are differently dependent on two-level
systems in dielectrics. Using NbTiN samples and introducing dipole
two-level systems by covering the samples with various SiO$_x$ layer
thicknesses we find that the logarithmic temperature dependent
increase in the resonance frequency scales with the layer thickness.
The frequency noise increases strongly as soon as a SiO$_x$ layer is
present and is, in contrast to the resonance frequency results,
thickness independent.

\begin{acknowledgments}
The authors thank A. Halvari, P. Kivinen and Y.~J.~Y. Lankwarden for
their contribution to the fabrication of the devices and J. N.
Hovenier for help with the experiment. The work was supported by
RadioNet (EU) under contract no. RII3-CT-2003-505818, the
Netherlands Organisation for Scientific Research (NWO) and
NanoFridge.
\end{acknowledgments}

\end{document}